\documentclass[aps,pre,amsmath,lengthcheck,superscriptaddress]{revtex4-2}

\usepackage{graphicx}\graphicspath{ {figures/} }
\usepackage{hyperref}
\hypersetup{colorlinks,allcolors=blue,breaklinks}

\newcommand{\be}{\begin{equation}}
\newcommand{\ee}{\end{equation}}
\newcommand{\bs}{\begin{split}}
\newcommand{\es}{\end{split}}
\newcommand{\ba}{\begin{align}}
\newcommand{\ea}{\end{align}}
\newcommand{\bi}{\begin{itemize}}
\newcommand{\ei}{\end{itemize}}

\newcommand{\la}{\left\langle}
\newcommand{\ra}{\right\rangle}
\newcommand{\pd}{\partial}

\newcommand{\bla}{bla\\bla\\bla\\bla\\bla}

\newcommand{\mb}[1]{\mbox{\boldmath$#1$}}
\newcommand{\mc}[1]{\mathcal{#1}}

\begin{document}

\title{Performance of near-optimal protocol in weakly driven processes}

\author{Pierre Naz\'e}
\email{pierre.naze@icen.ufpa.br}

\affiliation{\it Universidade Federal do Par\'a, Faculdade de F\'isica, ICEN,
Av. Augusto Correa, 1, Guam\'a, 66075-110, Bel\'em, Par\'a, Brazil}

\date{\today}

\begin{abstract}

A natural criticism of the optimal protocol of the irreversible work found for weakly driven processes is its experimental difficulty in being implementable due to its singular part. In this work, I explore the possibility of taking its continuous linear part as an acceptable near-optimal protocol. First, I prove that such a solution is the optimal protocol for non-singular admissible functions. I corroborate this result by observing successful comparisons with test protocols on six reasonable examples. Also, extending such analysis, I conclude that the error committed on this near-optimal protocol is considerable compared to the first-order singular approximation solution, except for sudden and slowly-varying processes. A conjecture is made about a general structure of a near-optimal protocol for systems under arbitrarily strong perturbations.

\end{abstract}

\maketitle

\section{Introduction}
\label{sec:introduction}

Optimization problems regarding the criticality of quantities in thermodynamic processes have been an intense area of research from its beginnings~\cite{carnot1960reflections,callen1991thermodynamics} until recent years~\cite{deffner2020thermodynamic,blaber2023optimal}. In particular, several works involving molecular thermal machines~\cite{seifert2012stochastic,schmiedl2007optimal,gomez2008optimal,kamizaki2022performance,engel2013optimal,loos2024universal} and quantum control~\cite{koch2019quantum,soriani2022three,soriani2022assessing} having been reported as a response to technical development recently achieved~\cite{ashkin1970acceleration,vandersypen2004nmr,arute2019quantum,hauke2020perspectives}.

Optimization problems are also very subject to the space of admissible functions of their solutions~\cite{gelfand2000calculus,kirk2004optimal}. In some cases, computational research may not be able to visualize completely some regions due to its proper limitations. The optimal protocols found in such a research are nothing more than near-optimal protocols~\cite{gingrich2016near,solon2018phase} when related to a bigger space of admissible functions. 

Another situation when this situation happens is when the admissible functions include generalized functions, such as Dirac deltas and their derivatives~\cite{naze2024analytical}. In particular, the experimental implementation of these cases can be challenging, although a proposition of how one can implement them has been proposed~\cite{naze2024analytical}. However, while such a procedure is not implementable, accessible near-optimal protocols, tangible to experimentalists, are of utmost importance.

In Ref.~\cite{naze2024analytical}, a universal optimal protocol for the irreversible work and its variance for isothermal and weak processes has been found, where the solution is split into a continuous linear part and a singular part, composed of Dirac deltas and their derivatives. In the face of the problem reported above, I propose as a near-optimal protocol the continuous linear part of the whole solution. 

As a first step, I prove that such a near-optimal protocol is the optimal protocol for non-singular admissible functions. Analyzing in this manner six reasonable examples, I corroborate such near-optimality. The analysis is extended to verify the optimality of the singular part and the performance of the near-optimal protocol compared to the first-order singular approximation expansion. The result is critical for regimes when the time of the process is equal to the timescale characteristic of the system, becoming better at sudden and slowly-varying processes. In the end, I conjecture a near-optimal protocol for systems under arbitrarily strong perturbations drivings based on Ref.~\cite{schmiedl2007optimal}.

\section{Preliminaries}
\label{sec:preliminaries}

I start by defining notations and developing the main concepts to be used in this work. This introductory technical section was based on Ref.~\cite{naze2024analytical}.

Consider a classical system with a Hamiltonian $\mc{H}(\mb{z}(\mb{z_0},t)),\lambda(t))$, where $\mb{z}(\mb{z_0},t)$ is a point in the phase space $\Gamma$ evolved from the initial point $\mb{z_0}$ until time $t$, with $\lambda(t)$ being a time-dependent external parameter. During a switching time $\tau$, the external parameter is changed from $\lambda_0$ to $\lambda_0+\delta\lambda$, with the system being in contact with a heat bath of temperature $\beta\equiv {(k_B T)}^{-1}$, where $k_B$ is Boltzmann's constant. The average work performed on the system during this interval of time is
\be
\langle W\rangle = \int_0^\tau \la\overline{\pd_{\lambda}\mc{H}}(t)\ra_0\dot{\lambda}(t)dt,
\label{eq:work}
\ee
where the generalized force $\la\overline{\pd_{\lambda}\mc{H}}\ra_0$ is calculated using the averaging $\overline{\cdot}$ over the stochastic path and the averaging $\langle\cdot\rangle_0$ over the initial canonical ensemble. The external parameter can be expressed as
\be
\lambda(t) = \lambda_0+g(t)\delta\lambda,
\label{eq:ExternalParameter}
\ee
where to satisfy the initial conditions of the external parameter the protocol $g(t)$ must satisfy the following boundary conditions $g(0)=0$ and $g(\tau)=1$. 

Linear-response theory expresses average quantities until the first-order of some perturbation parameter considering how this perturbation affects the observable to be averaged and the probability distribution \cite{kubo2012}. In this work, the parameter does not considerably change during the process, $|g(t)\delta\lambda/\lambda_0|\ll 1$, for all $t\in[0,\tau]$. In that manner, using such a framework, the average work can be split into the difference of free energy $\Delta F$ and irreversible work $W_{\rm irr}$~\cite{naze2022optimal}
\be
\Delta F = \delta\lambda\la\pd_{\lambda}\mc{H}\ra_0-\frac{\delta\lambda^2}{2}(\Psi_0(0)-\la\pd_{\lambda\lambda}^2\mc{H}\ra_0),
\ee  
\begin{equation}
\begin{split}
    W_{\rm irr} =& \frac{\delta\lambda^2}{2}\Psi(0)+\delta\lambda^2\int_0^\tau \dot{\Psi}_0(\tau-t)g(t)dt\\&-\frac{\delta\lambda^2}{2}\int_0^\tau\int_0^\tau \ddot{\Psi}(t-t')g(t)g(t')dt dt',
    \label{eq:wirr}
\end{split}
\end{equation}
where $\Psi_0(t)$ is the relaxation function, given by
\be
\Psi_0(t) = \beta\la\pd_\lambda\mc{H}(0)\overline{\pd_\lambda\mc{H}}(t)\ra_0-\mc{C},
\ee 
where the constant $\mc{C}$ is calculated to vanish the relaxation function for long times \cite{kubo2012}. To agree with the Second Law of Thermodynamics, the relaxation function must have its Fourier transform positive~\cite{naze2020,naze2024casimir}. Also, the relaxation timescale of the system
\be
\tau_R = \int_0^\infty \frac{\Psi_0(t)}{\Psi_0(0)}dt,
\ee
must be finite.

\begin{figure}
    \includegraphics[scale=0.85]{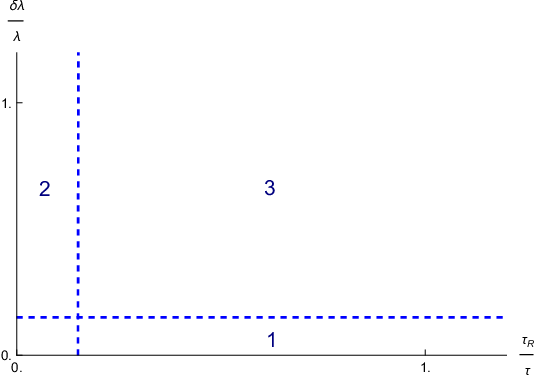}
    \caption{Diagram of nonequilibrium regions. Region 1: slowly-varying processes, Region 2: finite-time but weak processes and Region 3: arbitrarily far-from-equilibrium processes. Linear response theorem can describe regions 1 and 2.}
\label{fig:diagram}
\end{figure}

I establish at this point the regimes where linear-response theory is able to describe thermodynamic processes. Those regimes are determined by the relative strength of the driving with respect to the initial value of the protocol, $\delta\lambda/\lambda_0$, and the rate by which the process occurs with respect to the relaxation time of the system, $\tau_R/\tau$. See Fig.~\ref{fig:diagram} for a diagram depicting the regimes. In region 1, the so-called slowly-varying processes, the ratio $\delta\lambda/\lambda_0$ is arbitrary, while $\tau_R/\tau\ll 1$. By contrast, in region 2, the so-called finite-time and weak processes, the ratio $\delta\lambda/\lambda_0\ll 1$, while $\tau_R/\tau$ is arbitrary. In region 3, the so-called arbitrarily far-from-equilibrium processes, both ratios are arbitrary. Linear-response theory can only describe regions 1 and 2 \cite{naze2020}. In particular, the regime where $\tau_R/\tau\ll 1$ is called sudden processes.

Using the calculus of variations, we can derive the Euler-Lagrange equation that furnishes the optimal protocol $g^*(t)$ of the system that will minimize the irreversible work \cite{naze2022optimal}
\be
\int_0^\tau \ddot{\Psi}_0(t-t')g^*(t')dt' = \dot{\Psi}_0(\tau-t).
\label{eq:eleq}
\ee
In particular, the optimal irreversible work will be \cite{naze2022optimal}
\be
W_{\rm irr}^* = \frac{\delta\lambda^2}{2}\Psi_0(0)+\frac{\delta\lambda^2}{2}\int_0^\tau \dot{\Psi}_0(\tau-t)g^*(t)dt.
\ee
Also, the Euler-Lagrange equation \eqref{eq:eleq} furnishes also the optimal protocol that minimizes the variance of the work \cite{naze2023optimal}. In Ref.~\cite{naze2024analytical}, the following analytical solution was found
\be
g^*(t) = \frac{t+\tau_R}{\tau+2\tau_R}+\sum_{n=0}^{\infty}\frac{a_n (\delta^{(n)}(t)-\delta^{(n)}(\tau-t))}{\tau+2\tau_R},
\ee
where $a_n$ are terms connected to the relaxation function of the system. This work aims to respond to a natural criticism of such an analytical solution: due to its singular part, its experimental implementation can be a real challenge, so how can one overcome this problem? I present an analysis of the possibility of taking the continuous linear part as a reasonable near-optimal protocol. First, I prove that such a part is the optimal protocol of non-singular admissible functions. Second, I study six reasonable examples to evaluate the near-optimality of the linear continuous part, the optimality of the first-order singular part, and the performance of the near-optimal protocol with the first-order singular approximation expansion.

\section{Near-optimal protocol}

In this section, I am going to show that the continuous linear part of the whole solution
\be
g_{-1}^*(t)=\frac{t+\tau_R}{\tau+2\tau_R},
\label{eq:gclpart}
\ee
is the optimal protocol for non-singular admissible functions. In this way, it is a near-optimal protocol of the whole solution.

To prove this, consider that the near-optimal protocol $g(t)$, not satisfying anymore the Euler-Lagrange equation, can be expanded in shifted Chebyshev polynomials~\cite{bonancca2018minimal}, such that it can rewritten as a sum of a function $I$ with odd parity and other $J$ with even parity
\be
g(t)=I(t)+J(t),
\ee
with $I(t)=1-I(\tau-t)$ and $J(t)=J(\tau-t)$. In Ref.~\cite{bonancca2018minimal}, it has been shown that the basis that composes the odd parity leads to singular parts, except for the continuous linear part. Therefore, in a process of optimization, $I(t)$ would be the optimal protocol, with singular parts, and $J(t)=0$. To circumvent this problem, and work only with non-singular functions, I exclude this basis that leads to those singular parts and consider a $J(t)$ to be determined. The near-optimal protocol would be 
\be
g(t)=\frac{t+a}{\tau+2a}+J(t),
\label{eq:noprotocol1}
\ee
where $a$ is constant to be determined by the optimization problem. Observe that a general linear part should have two free parameters, but the odd parity condition in which the function has been expressed chooses one. Substituting Eq.~\eqref{eq:noprotocol1} in Eq.~\eqref{eq:wirr}, one will have
\be
\begin{split}
W_{\rm irr} &= W_{\rm irr}^{(l)}+\delta\lambda^2\int_0^\tau \dot{\Psi}_0(\tau-t)J(t)dt\\
&-\delta\lambda^2\int_0^\tau\int_0^\tau \ddot{\Psi}_0(t-t')\left(\frac{t+a}{\tau+2a}\right)J(t')dtdt'\\
&-\frac{\delta\lambda^2}{2}\int_0^\tau\int_0^\tau \ddot{\Psi}_0(t-t')J(t)J(t')dtdt',
\end{split}
\label{eq:wirr1}
\ee
where $W_{\rm irr}^{(l)}$ is the irreversible work associated with the continuous linear solution. Using now the even parity property of $J(t)$, one can show that
\be
\begin{split}
W_{\rm irr} &= W_{\rm irr}^{(l)}-\frac{\delta\lambda^2}{2}\int_0^\tau\int_0^\tau \ddot{\Psi}_0(t-t')J(t)J(t')dtdt',
\end{split}
\label{eq:wirr1}
\ee
whose minimal value is achieved only for $J(t)=0$ since $\Psi_0(t)$ has positive Fourier transform. Observe that the optimality condition did not impose any value to $a$. However, since it is predicted that the optimal protocol tends to $t/\tau$ for $\tau/\tau_R \rightarrow \infty$, the near-optimal protocol should have the same behavior. Therefore, for non-singular admissible functions, the optimal protocol is $g^*_{-1}(t)$, with optimal work
\be
\begin{split}
W_{\rm irr}^{(-1)} &= \frac{\delta\lambda^2}{2}\Psi_0(0)+\delta\lambda^2\int_0^\tau \dot{\Psi}_0(\tau-t)g^*_{-1}(t)dt\\
&-\frac{\delta\lambda^2}{2}\int_0^\tau\int_0^\tau \ddot{\Psi}_0(t-t')g^*_{-1}(t)g^*_{-1}(t')dtdt'.
\end{split}
\label{eq:wirr1}
\ee
\section{Analysis}

Considering the test protocols: a monotonic one
\be
g_{\rm test 1}(t) = \frac{t}{\tau},
\label{eq:gtest1}
\ee
and a non-monotonic one
\be
g_{\rm test 2}(t) = \frac{t}{\tau}+\sin{\left(\pi\frac{t}{\tau}\right)},
\label{eq:gtest2}
\ee
and the first-order singular approximation part
\be
g_{0}^*(t) = \frac{a_0(\delta(t)-\delta(\tau-t))}{\tau+2\tau_R},
\label{eq:gspart}
\ee
I will analyze six reasonable examples, where I will:
\begin{itemize}
    \item Corroborate $g_{-1}^*(t)$ as a near-optimal protocol compared to the test protocols;
    \item Verify the optimal convergence of $g_{0}^*(t)$;
    \item The performance of $g_{-1}^*(t)$ compared to the first-order singular approximation expansion $g_{-1}^*(t)+g_0^*(t)$.
\end{itemize}
To do so, I define
\be
\begin{split}
W_{\rm irr}^{(N)} &= \frac{\delta\lambda^2}{2}\Psi_0(0)+\delta\lambda^2\sum_{n=-1}^N\int_0^\tau \dot{\Psi}_0(\tau-t)g_n^*(t)dt\\
&-\frac{\delta\lambda^2}{2}\sum_{n=-1}^N\sum_{m=-1}^N\int_0^\tau\int_0^\tau \ddot{\Psi}_0(t-t')g_n^*(t)g_m^*(t')dtdt',
\end{split}
\ee
for $n\ge 0$. Also
\be
\begin{split}
W_{\rm irr}^{\rm (test 1)} &= \frac{\delta\lambda^2}{2}\Psi_0(0)+\delta\lambda^2\int_0^\tau \dot{\Psi}_0(\tau-t)g_{\rm test 1}(t)dt\\
&-\frac{\delta\lambda^2}{2}\int_0^\tau\int_0^\tau \ddot{\Psi}_0(t-t')g_{\rm test 1}(t)g_{\rm test 1}(t')dtdt',
\end{split}
\label{eq:wirrtest1}
\ee
\be
\begin{split}
W_{\rm irr}^{(\rm test 2)} &= \frac{\delta\lambda^2}{2}\Psi_0(0)+\delta\lambda^2\int_0^\tau \dot{\Psi}_0(\tau-t)g_{\rm test 2}(t)dt\\
&-\frac{\delta\lambda^2}{2}\int_0^\tau\int_0^\tau \ddot{\Psi}_0(t-t')g_{\rm test 2}(t)g_{\rm test 2}(t')dtdt'.
\end{split}
\label{eq:wirrtest1}
\ee
Observe that the analysis should be done with such a functional, since we deal with near-optimal protocol, and not the optimal one. The following error will be analyzed:
\be
\epsilon_{-1} = \left|\frac{W_{\rm irr}^{(0)}-W_{\rm irr}^{(-1)}}{W_{\rm irr}^{(0)}}\right|.
\label{eq:error}
\ee

\subsection{Examples}

I present now the examples and analysis for each case. All of them satisfy the positivity of the Fourier transform of the relaxation function~\cite{naze2024analytical}.

\subsubsection{Overdamped Brownian motion}

Consider a white noise overdamped Brownian motion subjected to a time-dependent harmonic potential, with the mass of the system equal to one, $\gamma$ as a damping coefficient, and $\omega_0$ as the natural frequency of the potential. The relaxation function for both moving laser and stiffening traps~\cite{naze2022optimal} are given by
\be
\Psi_1(t)=\Psi_0(0) \exp{\left(-\frac{|t|}{\tau_R}\right)},
\ee
where $\tau_R$ is the relaxation timescale of each case. In this case, the optimal protocol under weak drivings is just the linear protocol~\cite{schmiedl2007optimal,naze2022optimal}. Therefore, a discussion about the near-optimality, optimal convergence and performance related to the near-optimal protocols is not necessary.

\begin{figure}[t]
    \centering
    \includegraphics[scale=0.85]{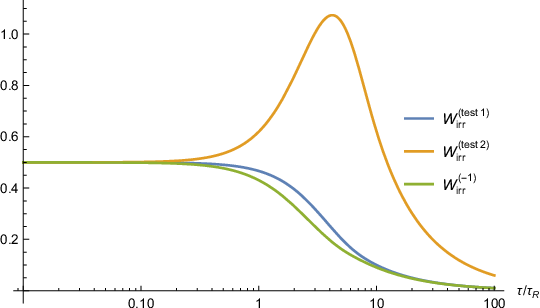}
    \caption{Corroboration of near-optimality of the continuous linear part \eqref{eq:gclpart} compared to test protocols \eqref{eq:gtest1} and \eqref{eq:gtest2} for $\Psi_2(t)$. It was used $\gamma=1$, $\omega_0=1$ and work units of $\delta\lambda^2\Psi_0(0)$.}
    \label{fig:psi21}
\end{figure}

\begin{figure}[t]
    \centering
    \includegraphics[scale=0.85]{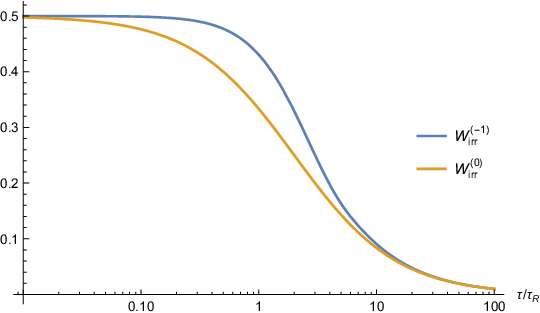}
    \caption{Corroboration of optimal convergence for the irreversible work of the first-order singular approximation expansion with \eqref{eq:gspart} compared to continuous linear part \eqref{eq:gclpart} for $\Psi_2(t)$. It was used $\gamma=1$, $\omega_0=1$ and work units of $\delta\lambda^2\Psi_0(0)$.}
    \label{fig:psi22}
\end{figure}

\begin{figure}[t]
    \centering
    \includegraphics[scale=0.85]{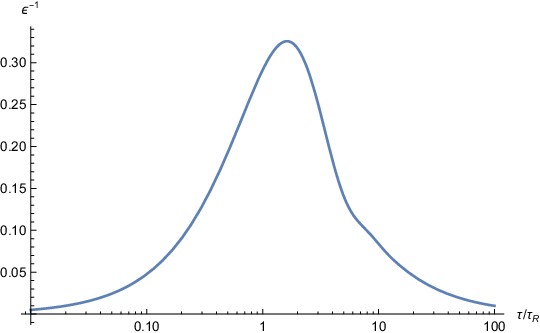}
    \caption{Error \eqref{eq:error} comparing the irreversible work of the continuous linear part with the first-order singular approximation expansion for $\Psi_2(t)$. It achieves its peak around $\tau\approx\tau_R$, becoming significantly better at sudden and slowly-varying processes. It was used $\gamma=1$, $\omega_0=1$ and work units $\delta\lambda^2\Psi_0(0)$.}
    \label{fig:psi23}
\end{figure}

\subsubsection{Underdamped Brownian motion:\\ moving laser trap}

Consider a white noise underdamped Brownian motion subjected to a time-dependent harmonic potential, with $m$ as the mass of the particle, $\gamma$ as a damping coefficient, and $\omega_0$ as the natural frequency of the potential. The relaxation function for moving laser trap~\cite{naze2022optimal} is given by
\be
\Psi_2(t)=\Psi_0(0) \exp{\left(-\frac{\gamma}{2}|t|\right)}\left(\cos{\omega t}+\frac{\gamma}{2\omega}\sin{\omega |t|}\right),
\ee
where $\omega=\sqrt{\omega_0^2-\gamma^2/4}$ is a positive number. The relaxation time is given by $\tau_R=\gamma/\omega_0^2$~\cite{naze2022optimal}.

Using $\gamma=1$ and $\omega_0=1$, in Fig.~\ref{fig:psi21}, we observe that the continuous linear part has the minimal value compared to the test protocols. In Fig.~\ref{fig:psi22}, the addition of Dirac deltas becomes the solution more optimal than just using the linear continuous part. It is important to remember that the optimal solution in this case was already achieved by using such an addition~\cite{naze2022optimal}. In Fig.~\ref{fig:psi23}, the error committed to achieving the optimal work achieves a peak around $30\%$ around $\tau\approx \tau_R$, but it becomes less than $5\%$ for sudden and slowly-varying processes. These behaviors are explained by the participation of Dirac deltas in each regime considered~\cite{naze2024analytical}.

\subsubsection{Underdamped Brownian motion:\\ stiffening laser trap}

For the same system of the previous example, but considering the stiffening trap case, the relaxation function is~\cite{naze2020}
\begin{multline}
\Psi_3(t)=\Psi_0(0) \exp{\left(-\gamma|t|\right)}\left[\frac{2\omega_0^2}{\omega^2}\right.\\
\left.+\left(\frac{\omega^2-2\omega_0^2}{\omega^2}\right)\cos{\omega t}+\frac{\gamma}{\omega}\sin{\omega |t|}\right],
\end{multline}
where $\omega=\sqrt{4\omega_0^2-\gamma^2}$ is a positive number. The relaxation time is $\tau_R=(\gamma^2+\omega_0^2)/(2\gamma\omega_0^2)$ .

\begin{figure}
    \centering
    \includegraphics[scale=0.85]{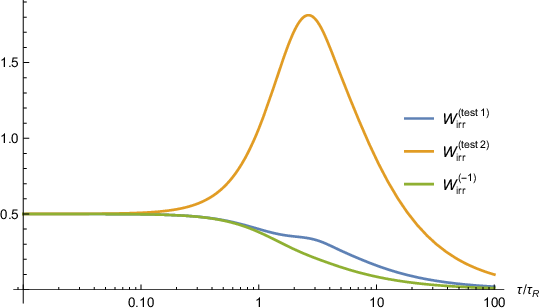}
    \caption{Corroboration of near-optimality of the continuous linear part \eqref{eq:gclpart} compared to test protocols \eqref{eq:gtest1} and \eqref{eq:gtest2} for $\Psi_3(t)$. It was used $\gamma=1$, $\omega_0=1$ and work units of $\delta\lambda^2\Psi_0(0)$.}
    \label{fig:psi31}
\end{figure}

\begin{figure}
    \centering
    \includegraphics[scale=0.85]{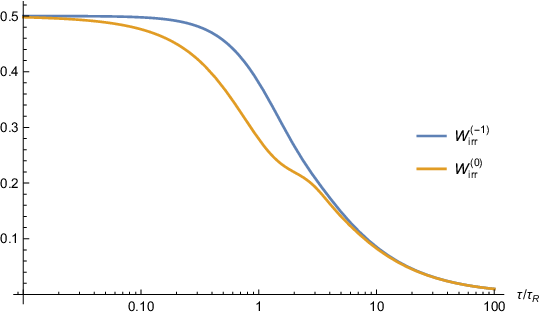}
    \caption{Corroboration of optimal convergence for the irreversible work of the first-order singular approximation expansion with \eqref{eq:gspart} compared to continuous linear part \eqref{eq:gclpart} for $\Psi_3(t)$. It was used $\gamma=1$, $\omega_0=1$ and work units of $\delta\lambda^2\Psi_0(0)$.}
    \label{fig:psi32}
\end{figure}

\begin{figure}
    \centering
    \includegraphics[scale=0.85]{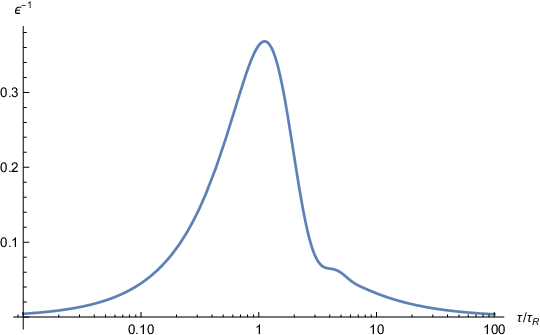}
    \caption{Error \eqref{eq:error} comparing the irreversible work of the continuous linear part with the first-order singular approximation expansion for $\Psi_3(t)$. It achieves its peak around $\tau=\tau_R$, becoming significantly better at sudden and slowly-varying processes. It was used $\gamma=1$, $\omega_0=1$ and work units $\delta\lambda^2\Psi_0(0)$.}
    \label{fig:psi33}
\end{figure}

Using $\gamma=1$ and $\omega_0=1$, in Fig.~\ref{fig:psi31}, we observe that the continuous linear part has the minimal value compared to the test protocols. In Fig.~\ref{fig:psi32}, the addition of Dirac deltas becomes the solution more optimal than just using the linear continuous part. It is important to remember that the optimal solution in this case is not achieved by using such an addition. In Fig.~\ref{fig:psi33}, the error committed to achieving the better work achieves a peak around $40\%$ around $\tau\approx \tau_R$, but it becomes less than $5\%$ for sudden and slowly-varying processes. As before, these behaviors are explained by the participation of Dirac deltas in each regime considered~\cite{naze2024analytical}.

\subsubsection{Bessel relaxation function}

The Bessel relaxation function is given by
\be
\Psi_4(t)=\Psi_0(0) J_0\left(\frac{t}{\tau_R}\right),
\ee
where $J_0$ is the Bessel function of the first kind with $\nu=0$ and $\tau_R$ is its relaxation timescale. It satisfies the criteria for compatibility with the Second Law of Thermodynamics. Such relaxation function can model the Ising chain subjected to a time-dependent magnetic field and evolving in time at equilibrium accordingly to Glauber-Ising dynamics~\cite{glauber1963time}.

\begin{figure}
    \centering
    \includegraphics[scale=0.85]{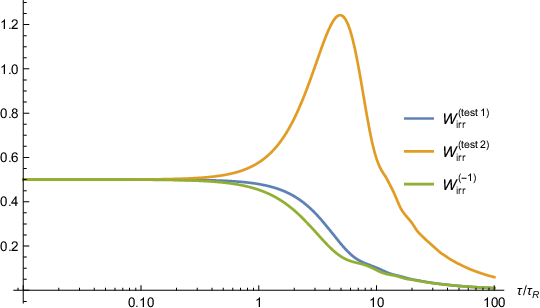}
    \caption{Corroboration of near-optimality of the continuous linear part \eqref{eq:gclpart} compared to test protocols \eqref{eq:gtest1} and \eqref{eq:gtest2} for $\Psi_4(t)$. It was used $\tau_R=1$ and work units of $\delta\lambda^2\Psi_0(0)$.}
    \label{fig:psi41}
\end{figure}

\begin{figure}
    \centering
    \includegraphics[scale=0.85]{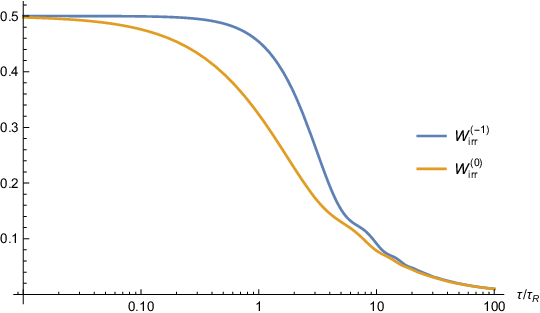}
    \caption{Corroboration of optimal convergence for the irreversible work of the first-order singular approximation expansion with \eqref{eq:gspart} compared to continuous linear part \eqref{eq:gclpart} for $\Psi_4(t)$. It was used $\tau_R=1$ and work units of $\delta\lambda^2\Psi_0(0)$.}
    \label{fig:psi42}
\end{figure}

\begin{figure}
    \centering
    \includegraphics[scale=0.85]{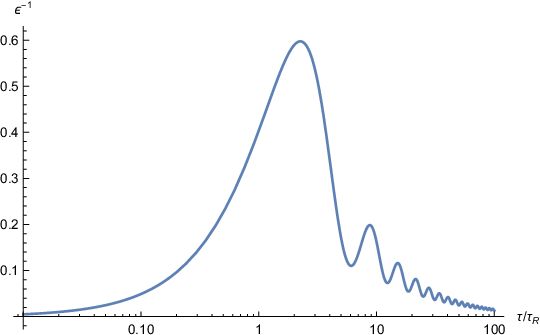}
    \caption{Error \eqref{eq:error} comparing the irreversible work of the continuous linear part with the first-order singular approximation expansion for $\Psi_4(t)$. It achieves its peak around $\tau\approx\tau_R$, becoming significantly better at sudden and slowly-varying processes. It was used $\tau_R=1$ and work units $\delta\lambda^2\Psi_0(0)$.}
    \label{fig:psi43}
\end{figure}

Using $\tau_R=1$, in Fig.~\ref{fig:psi41}, we observe that the continuous linear part has the minimal value compared to the test protocols. In Fig.~\ref{fig:psi42}, the addition of Dirac deltas becomes the solution more optimal than just using the linear continuous part. It is important to remember that the optimal solution in this case is not achieved by using such an addition. In Fig.~\ref{fig:psi43}, the error committed to achieving the better work achieves a peak around $60\%$ around $\tau\approx \tau_R$, but it becomes less than $5\%$ for sudden and slowly-varying processes. As before, these behaviors are explained by the participation of Dirac deltas in each regime considered~\cite{naze2024analytical}.

\subsubsection{Gaussian relaxation function}

A relaxation function that satisfies the criteria of compatibility with the Second Law of Thermodynamics~\cite{naze2020} is the Gaussian relaxation function
\be
\Psi_5(t)=\Psi_0(0) \exp\left(-\frac{\pi}{4}\left(\frac{t}{\tau_R}\right)^2\right),
\ee
where $\tau_R$ is the relaxation timescale of the system. 

\begin{figure}
    \centering
    \includegraphics[scale=0.85]{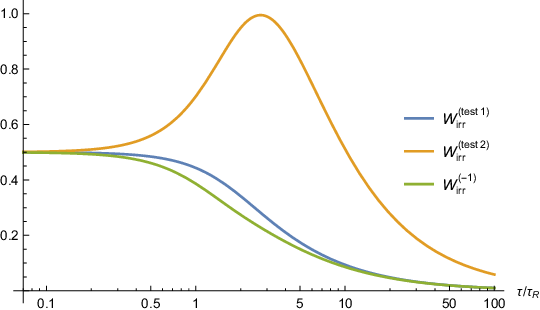}
    \caption{Corroboration of near-optimality of the continuous linear part \eqref{eq:gclpart} compared to test protocols \eqref{eq:gtest1} and \eqref{eq:gtest2} for $\Psi_5(t)$. It was used $\tau_R=1$ and work units of $\delta\lambda^2\Psi_0(0)$.}
    \label{fig:psi51}
\end{figure}

\begin{figure}
    \centering
    \includegraphics[scale=0.85]{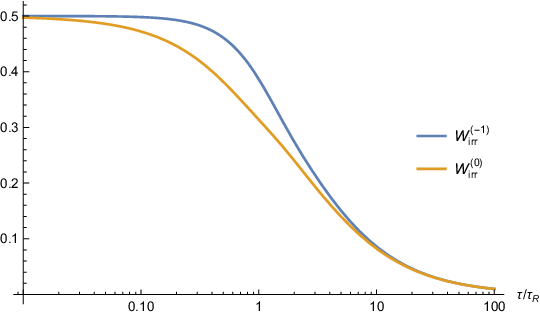}
    \caption{Corroboration of optimal convergence for the irreversible work of the first-order singular approximation expansion with \eqref{eq:gspart} compared to continuous linear part \eqref{eq:gclpart} for $\Psi_5(t)$. It was used $\tau_R=1$ and work units of $\delta\lambda^2\Psi_0(0)$.}
    \label{fig:psi52}
\end{figure}

\begin{figure}
    \centering
    \includegraphics[scale=0.85]{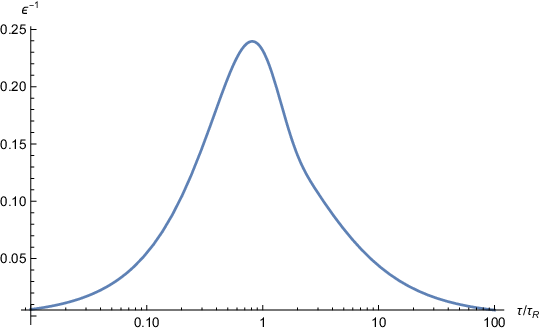}
    \caption{Error \eqref{eq:error} comparing the irreversible work of the continuous linear part with the first-order singular approximation expansion for $\Psi_5(t)$. It achieves its peak around $\tau\approx\tau_R$, becoming significantly better at sudden and slowly-varying processes. It was used $\tau_R=1$ and work units $\delta\lambda^2\Psi_0(0)$.}
    \label{fig:psi53}
\end{figure}

Using $\tau_R=1$, in Fig.~\ref{fig:psi51}, we observe that the continuous linear part has the minimal value compared to the test protocols. In Fig.~\ref{fig:psi52}, the addition of Dirac deltas becomes the solution more optimal than just using the linear continuous part. It is important to remember that the optimal solution in this case is not achieved by using such an addition. In Fig.~\ref{fig:psi53}, the error committed to achieving the better work achieves a peak around $25\%$ around $\tau\approx \tau_R$, but it becomes less than $5\%$ for sudden and slowly-varying processes. As before, these behaviors are explained by the participation of Dirac deltas in each regime considered~\cite{naze2024analytical}.

\begin{figure}
    \centering
    \includegraphics[scale=0.85]{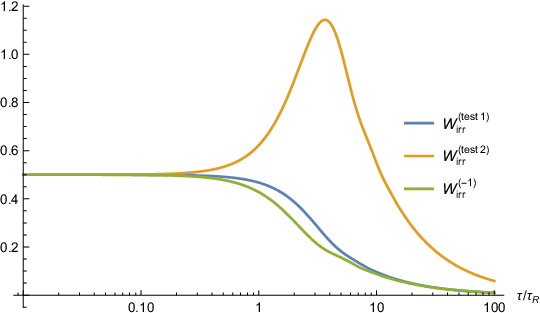}
    \caption{Corroboration of near-optimality of the continuous linear part \eqref{eq:gclpart} compared to test protocols \eqref{eq:gtest1} and \eqref{eq:gtest2} for $\Psi_6(t)$. It was used $\tau_R=1$ and work units of $\delta\lambda^2\Psi_0(0)$.}
    \label{fig:psi61}
\end{figure}

\begin{figure}
    \centering
    \includegraphics[scale=0.85]{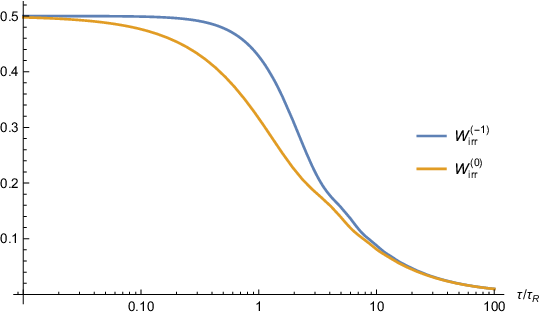}
    \caption{Corroboration of optimal convergence for the irreversible work of the first-order singular approximation expansion with \eqref{eq:gspart} compared to continuous linear part \eqref{eq:gclpart} for $\Psi_6(t)$. It was used $\tau_R=1$ and work units of $\delta\lambda^2\Psi_0(0)$.}
    \label{fig:psi62}
\end{figure}

\begin{figure}
    \centering
    \includegraphics[scale=0.85]{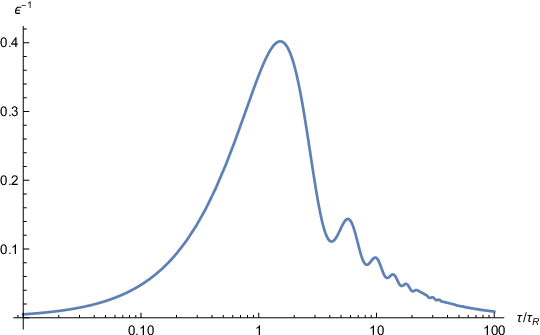}
    \caption{Error \eqref{eq:error} comparing the irreversible work of the continuous linear part with the first-order singular approximation expansion for $\Psi_6(t)$. It achieves its peak around $\tau\approx\tau_R$, becoming significantly better at sudden and slowly-varying processes. It was used $\tau_R=1$ and work units $\delta\lambda^2\Psi_0(0)$.}
    \label{fig:psi63}
\end{figure}

\subsubsection{Sinc relaxation function}

In Ref.~\cite{naze2023adiabatic}, when we apply the method of time average in a thermally isolated system performing an adiabatic process, we produce a new one performing an isothermal one with a typical relaxation time. In particular, for thermally isolated systems that have a relaxation function equal to
\be
\Psi_0(t)=\Psi_0(0) \cos{(\omega t)},
\ee
will have for time-averaged relaxation function 
\be
\Psi_6(t)=\Psi_0(0)\,\text{sinc}\left(\frac{\pi}{2} \frac{t}{\tau_R}\right),
\ee
where $\tau_R$ is the relaxation timescale of the system. 

Using $\tau_R=1$, in Fig.~\ref{fig:psi61}, we observe that the continuous linear part has the minimal value compared to the test protocols. In Fig.~\ref{fig:psi62}, the addition of Dirac deltas becomes the solution more optimal than just using the linear continuous part. It is important to remember that the optimal solution in this case is not achieved by using such an addition. In Fig.~\ref{fig:psi63}, the error committed to achieving the better work achieves a peak around $40\%$ around $\tau\approx \tau_R$, but it becomes less than $5\%$ for sudden and slowly-varying processes. As before, these behaviors are explained by the participation of Dirac deltas in each regime considered~\cite{naze2024analytical}.

\section{Conjecture}

In Ref.~\cite{schmiedl2007optimal} the optimal protocol for a white noise overdamped Brownian motion subjected to a stiffening harmonic trap is calculated analytically, given by
\begin{widetext}
\be
g^*(t) = \frac{1}{\delta \lambda}(\lambda^*(t)-\lambda_0),
\ee
with
\be
\lambda^*(t)= \lambda_0\left(\frac{1}{(1+c^* t)^2}-\frac{2\tau_Rc^*}{(1+c^* t)}\right),
\ee
\be
c^*=\frac{1}{\tau}\left(\frac{-2-\tau/\delta\tau_R-\tau/\tau_R+\sqrt{1+\tau/\tau_R+\tau^2/(4\tau_R\delta\tau_R)+\tau^2/(4\tau_R^2)}}{4+\tau/\delta\tau_R+\tau/\tau_R}\right),
\ee
\end{widetext}
where $\delta\tau_R$ is the relaxation time calculated with the variation of the parameter at the end of the process. Its approximation in weakly driven regimes furnishes the near-optimal protocol calculated in this work, which structure holds as a near-optimal protocol for all kinds of Hamiltonian systems. I conjecture that the above expression gives a general structure for the near-optimal protocol for higher-order perturbations for all Hamiltonian systems. For a general structure for optimal protocols, which includes singular parts, a detailed study of an exactly solvable example that furnishes such characteristics must be made for a similar conjecture.

\section{Final remarks}
\label{sec:final} 

In this work, I analyzed the continuous linear part approximation of the analytical optimal protocol for weak processes as a reasonable near-optimal protocol. First, I show that such a function is the best protocol for non-singular admissible functions. Considering six relevant examples, I verified a corroboration of such near-optimality and that the addition of Dirac deltas leads to lesser works, with significant error involved compared to the continuous linear case for regimes where $\tau\approx\tau_R$. The result is better for sudden and slowly-varying processes, with error less than 5\%. Adding higher-order singular parts, involving derivatives of Dirac deltas, becomes increasingly difficult given the double integral involved in the calculation of work. It is important to remark that the error found in this work is a lower bound since the comparisons were made between two different approximations. Finally, a conjecture is made about a general structure of the near-optimal protocol for non-singular admissible functions of systems under arbitrarily strong perturbation drivings.

\bibliography{PNOP}

\end{document}